\documentclass[doublecol]{epl2}

\usepackage{amsmath}
\usepackage{tabularx}
\usepackage{graphicx}
\usepackage{color}
\usepackage{times}

\graphicspath{{data/}}

\newcommand{\be}{\begin{equation}}
\newcommand{\ee}{\end{equation}}
\newcommand{\bea}{\begin{eqnarray}}
\newcommand{\eea}{\end{eqnarray}}

\newcommand{\beq}{\begin{equation}}
\newcommand{\eeq}{\end{equation}}
\newcommand{\beqn}{\begin{eqnarray}}
\newcommand{\eeqn}{\end{eqnarray}}

\title{High-precision simulation of the
height distribution for the KPZ equation}

\author{Alexander K. Hartmann\inst{1,2} \and Pierre Le Doussal\inst{3}
\and Satya N. Majumdar\inst{1} \and Alberto Rosso \inst{1} \and 
Gregory Schehr\inst{2}}
\shortauthor{A.K. Hartmann \etal}
\institute{
\inst{1} Institut f\"ur Physik, Universit\"at Oldenburg, 
	     26111 Oldenburg, Germany\\
\inst{2} LPTMS, CNRS, Univ. Paris-Sud, Universit\'e Paris-Saclay, 
91405 Orsay, France\\
\inst{3} CNRS-Laboratoire de Physique Th\'eorique de 
l'Ecole Normale Sup\'erieure, 24 rue Lhomond, 75231 Paris Cedex, France
}
\pacs{05.10.Ln}{Monte Carlo methods}
\pacs{75.10.Nr}{Spin-glass and other random models}
\pacs{05.20.-y}{Classical statistical mechanics}

\abstract{
The one-point distribution of the height for the continuum Kardar-Parisi-Zhang (KPZ)
equation is determined numerically using the mapping to the directed polymer in
a random potential at high temperature. Using an importance sampling approach, the distribution is
obtained over a large range of values, down to a probability density 
as small as $10^{-1000}$ in the tails. Both short and long times are investigated and compared with recent analytical predictions for the large-deviation forms of the probability of rare fluctuations.
At short times the agreement with the analytical expression is spectacular. We observe
that the far left and right tails, with exponents $5/2$ and $3/2$ respectively, 
are preserved until large time. We present some evidence for the predicted 
non-trivial crossover in the left tail from the $5/2$ tail exponent to the cubic tail 
of Tracy-Widom, although the details of the full scaling form remains beyond reach.
}

\begin{document}
\maketitle

\section{Introduction}
\label{sec:intro}

The $1+1$ dimensional Kardar-Parisi-Zhang (KPZ) equation
describes the non-linear stochastic growth of an interface \cite{kardar1986}. 
It is also relevant in a wide variety of physical models ranging from directed polymers in random media 
\cite{kardar1986,halpin1995kinetic,johansson2000shape,prahofer2000universal,prahofer2002scale,calabrese2010free,dotsenko2010bethe} to asymmetric exclusion process
models for the transport of interacting particles \cite{ferrari2006scaling,de2011large,kriecherbauer2010pedestrian,schutz2001exactly}
and has a number of experimental realizations \cite{takeuchi2010universal,takeuchi2011growing,miettinen2005experimental}. 
The interface is described by a field
$h(x,t)$ that denotes its height at the position  $x$ and at time $t$. The KPZ equation of motion is 
\be
\label{eq:KPZ}
\partial_t h = \nu \, \partial_x^2 h + \frac{\lambda_0}{2}\, (\partial_x h)^2 + \sqrt{D} \, \xi(x,t) \;,
\ee
where 
$\nu > 0$ gives the strength of the diffusive relaxation, $\lambda_0 > 0$ is the coefficient of the 
non-linearity and 
$\xi(x,t)$ is a Gaussian white noise with zero mean and 
$\langle \xi(x,t) \xi(x',t')\rangle = \delta(x-x')\delta(t-t')$.   From dimensional analysis it 
is  natural to introduce the following characteristic scales of space $x^*=(2 \nu)^3/(D \lambda_0^2)$, 
time $t^*=2(2 \nu)^5/(D^2 \lambda_0^4)$ and height
$h^*=\frac{2 \nu}{\lambda_0}$. For simplicity in the following we will work in rescaled units: $x/x^* \to x$, $t/t^* \to t$, $h/h^* \to h$. 
At large times $t \gg 1$ it is known that, due to the non-linearity, 
  the interface moves with a finite deterministic velocity $v_\infty$ 
which depends on the initial condition.

In the last decades tremendous progress has been achieved in obtaining exact results on the statistics of the height fluctuations \cite{huse1985huse, halpin1995kinetic, krug1997origins, corwin2012kardar}, e.g. of the centered height at one space point defined as
$H(t)=h(x=0,t)-v_\infty t+ \frac{1}{2} \ln t$. In particular the best studied case corresponds to a narrow wedge intial condition $h(x,t=0)= - |x|/\delta$ with $\delta \ll 1$ which gives rise at late times to the experimentally relevant curved or {\em droplet}  profile. In this case the fluctuations of $H$ can be expressed, for any time $t$, in terms of a Fredholm determinant \cite{sasamoto2010one, calabrese2010free, dotsenko2010bethe, amir2011probability}. Despite this exact result, since the Fredholm determinant is a complicated mathematical object,  it remains very challenging to obtain useful explicit information about the statistics of $H$ at a given time $t$. It is known that at short time, $t\ll 1$, the non-linear term in Eq.~(\ref{eq:KPZ}) is less important compared to the linear Laplacian term. In this limit the typical fluctuations of $H$ are well described by the Edwards-Wilkinson equation (i.e. Eq.~(\ref{eq:KPZ}) with $\lambda_0=0$). Hence in the short time limit the typical fluctuations of $H$ are of order $\sim t^{1/4}$ and  Gaussian.
On the other hand at large time, $t\gg 1$, the typical fluctuations of order $\sim t^{1/3}$ are described by the Tracy-Widom (TW) distribution associated to the typical fluctuations of the largest eigenvalue of random matrices belonging to the Gaussian Unitary Ensemble (GUE) \cite{tracy1994level}. The TW distribution  has been observed experimentally in nematic liquid crystals which exhibits KPZ growth laws \cite{takeuchi2010universal,takeuchi2011growing}.
\begin{figure}[ht]
\includegraphics[width=0.95\linewidth]{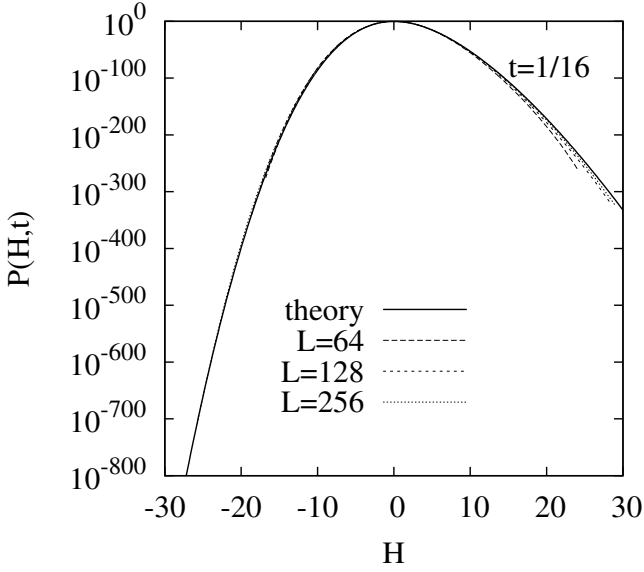}
\caption{Distribution of $P(H,t)$ for a short time $t=1/16$ for three
different lengths $L=64$, $L=128$ and $L=256$. The solid line indicates the analytical result in Eq. (\ref{eq:ST}) obtained in Ref. \cite{leDoussal2016short}. The agreement between numerical and analytical results is extremely good (on the left tail, down to values of the order $10^{-800}$).}\label{Fig1}
\end{figure}
More recently there has been an increasing interest in computing probability of rare fluctuations of $H$ away from its typical values. This large-deviation problem can be addressed both for short and large times.
The question of whether and how the tails evolve with time is important for many models in the KPZ class. Here we explore this issue numerically for the KPZ equation itself, and compare with recent analytical predictions.
In particular thanks to a short time expansion of the exact Fredholm determinant formula an explicit form for  the short time distribution $P(H,t)$, with $t \ll 1$, has been obtained \cite{leDoussal2016short}. It takes a large-deviation form:

\begin{equation}
P(H,t) \sim c(t)e^{-\frac 1 {\sqrt{t}}\phi_{\text{short}}(H)} \label{eq:ST}
\end{equation}
where $c(t)$ is a time dependent normalisation constant. The exact form of $\phi_{\text{short}}(H)$ is given in \cite{leDoussal2016short}, its asymptotic behavior, which can also be obtained using weak noise theory  \cite{kamenev2016short} reads \cite{kamenev2016short, leDoussal2016short}:

\begin{eqnarray}
\phi_{\rm short}(H) \simeq
\begin{cases}
\dfrac{4}{15 \pi} |H|^{5/2} \quad , \quad H \to - \infty  \label{asympt1} \\
\\
\dfrac{H^2}{\sqrt{2 \pi}} \quad , \quad \quad  \quad \quad |H| \ll 1 \label{asympt2} \\
\\
\dfrac{4}{3} H^{3/2} \quad , \quad  \quad \quad H \to + \infty \;.
\label{asympt3} 
\end{cases}
\end{eqnarray}
As expected, the typical fluctuations around $H=0$ are Gaussian, but the tails are asymmetric. In particular the right tail, $ H \to + \infty$, coincides exactly with the TW tail, while the left tail is characterized by a different $\frac{4}{15 \pi}  |H|^{5/2}$ behaviour, different from the $\frac{1}{12}|H|^{3}$ of the TW distribution. The tail behaviours  $\propto |H|^{5/2}$ (left) and $\propto H^{3/2}$ (right)  seem to be quite robust with respect to different initial conditions: indeed it has been obtained at short time also for flat as well as stationary initial conditions albeit with different prefactors \cite{kolokolov2007optimal,meerson2016large, kamenev2016short, krajenbrink2017exact}. In addition the central part of the distribution depends on the initial condition.

Exact results for the large deviations have also been obtained at long time, $t \gg 1$, for the droplet initial condition. In particular $P(H,t)$  displays three different regimes \cite{le2016large}:

\begin{eqnarray}\label{main_results}
\hspace*{-0cm}P(H,t) \sim
\begin{cases}
& e^{-t^2\,\Phi_-(H/t)} \quad , \quad  H \sim {\cal O}(t)  < 0 \; \quad  \label{1} \\
&\\
&\dfrac{1}{t^{1/3}} f_2\left[ \dfrac{H}{t^{1/3}} \right] \;, \; \hspace*{0.3cm} 
H \sim {\cal O}(t^{1/3}) \quad  \label{2} \\
&\\
& e^{-t \,\Phi_+(H/t)} \quad , \quad  H \sim {\cal O}(t) > 0 \quad  \label{3} \;,
\end{cases}
\end{eqnarray}
where $f_2(z)$ is the  GUE TW distribution. The tails have also been computed explicitly. The right tail rate function \cite{le2016large}

\begin{equation}
\Phi_+(z)=\frac{4}{3} z^{3/2} \label{eq:phi+}
\end{equation}
coincides exactly with the TW tail as already observed in the short time regime. The left tail rate function
was predicted in \cite{sasorov2017} to be

\begin{equation}
\Phi_-(z) =  
\frac{4}{15\pi^6}(1-\pi^2z)^{5/2}-\frac{4}{15\pi^6}+\frac{2}{3\pi^4}z
-\frac{1}{2\pi^2}z^2. \label{eq:phi_m}
\end{equation}
Note that Eq.~(\ref{eq:phi_m}) exhibits a crossover between two distinct tail behaviours of $P(H,t)$ for large negative $H$:  when $z=\frac{H}{t}~\to~0$  one has $\Phi_-(z) \simeq |z|^3/12$ such that from the first line of Eq.~(\ref{2}) one recovers the left tail of the TW distribution, i.e. $P(H,t) \sim e^{-|H|^3/(12t)}$. On the other hand when $z=\frac{H}{t}~\to~-\infty$ one has $\Phi_-(z) \simeq \frac{4}{15 \pi} |z|^{5/2}$, which coincides with the left tail of the short time large deviation given in the first line of Eq.~(\ref{asympt2}), i.e. $P(H,t) \sim e^{-\frac{4}{15 \pi} |H|^{5/2}/\sqrt{t}}$. 

For intermediate time, $t\sim 1$, only the cumbersome Fredholm determinant formula is available and no explicit information is known for large fluctuations of $H$.
Indeed numerical results focused only on the typical fluctuations  \cite{calabrese2010free, leDoussal2016short} as the study of the tails  requires a huge number of samples.

 In this paper we use importance sampling techniques and study numerically the full large deviations of $H$ both at short and intermediate time. This allows us to explore the tail statistics with an unprecedented precision of the order of $10^{-1000}$.
For short time our results perfectly agree with the theoretical prediction in Eq.~(\ref{asympt2}) (see Fig. \ref{Fig1}) and the asymptotic behaviour of the tails is clearly seen in Fig.~\ref{fig:P_H_0625}, both for the left tail (left panel) and the right tail (right panel). For the intermediate times our results are consistent with the following scenario (see Fig. \ref{fig:P_H_compare}) : (i) the right tail $P(H,t)~\sim~\exp(-\frac{4}{3} H^{3/2}/\sqrt{t})$ remains valid at all times (ii) the left tail is well described by  $\Phi_-(z)$ for large negative $z=H/t$, i.e.
the $5/2$ exponent remains valid at all times
(iii) the small $z$ behaviour of  $\Phi_-(z)$ and the typical fluctuations of $H$ have not yet reached the TW limiting behaviour. Larger time than the ones accessible in our simulations are needed to observe the large time TW behaviour and to fully confirm the form \eqref{eq:phi_m} for $\Phi_-(z)$.

\section{Model and Algorithm}

\begin{figure}[!htb]
\begin{center}
\includegraphics[width=0.8\linewidth]{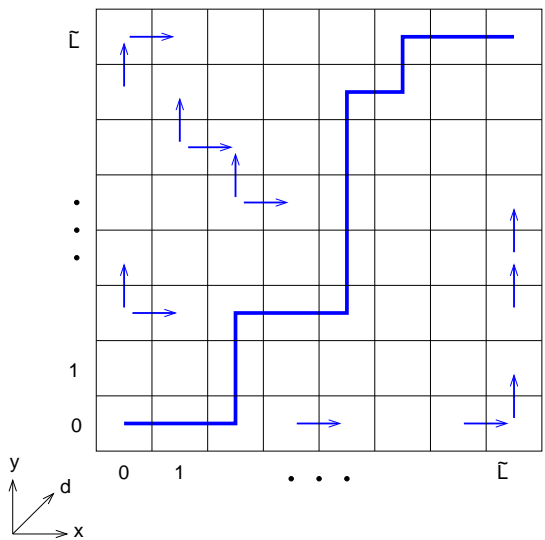}
\end{center}
\caption{Setup of the lattice with examples of possible bonds of the
polymer (small arrows) and one example of a polymer (thick line). 
All polymers start at $(0,0)$ and end in $(\tilde L,\tilde L)$
and therefore consist of $L=2\tilde L$ bonds.
\label{fig:lattice}}
\end{figure}

There is a standard mapping between the height in the KPZ and the free energy of a directed polymer 
at high temperature
embedded in a 1+1 random potential \cite{calabrese2010free,bustingorry2010universal}.
For a polymer of size $L=2\tilde L$ bonds,
the realisation of the disordered potential
 is given by a two dimensional lattice of 
$(\tilde L+1)\times (\tilde L+1)$
random numbers $V[x][y]$ ($x,y=0,1,\ldots,\tilde L$)
drawn from a Gaussian distribution $N(0,1)$, i.e., with mean 0
and variance 1. We consider all polymers which start at $(0,0)$ and
end at $(\tilde L,\tilde L)$, such that the polymer continues onto
neighboring sites of the lattice given that
the ``diagonal''  direction $d=x+y$ increases by one. The geometric
setup is shown in Fig.\ \ref{fig:lattice}.

A polymer visiting a set $P$ of sites has an energy

\begin{equation}
E_V(P)=\sum_{(x,y)\in P} V[x][y]\,.
\end{equation}
We are interested in the canonical ensemble, where each polymer
in the disorder landscape $V\equiv\{V[x][y]\}$ 
is connected to a heat bath with temperature $T$ and exhibts a 
Boltzmann weight

\begin{equation}
w_V(P)=e^{-E_V(P)/T}\,. \label{eq:polymer:Boltzmann}
\end{equation}
Therefore, for a given disorder realisation $V$ 
the partition function $Z(V)$ is given by

\begin{equation}
Z(V)=\sum_{P} w_V(P)\,,
\label{eq12}
\end{equation}
where the sum runs over all possible polymers with requirements
as explained above. Due to the requirement that the polymer extends
only in increasing diagonal value $d$, the partition function can
be calculated recursively using :

\begin{eqnarray}
Z[x][y] & = & (Z[x-1][y]+Z[x][y-1])e^{-V[x][y]/T} 
\end{eqnarray} 
where $Z[x][y]$ is the partition function of the polymer starting at $(0,0)$ and ending at $(x,y)$. Thus
the partition function defined in Eq.~(\ref{eq12}) is given by
$Z(V)=Z[\tilde L][\tilde L]$ and requires $O(\tilde L^2)$ steps to be computed.
The mapping between the free energy of the directed polymer at temperature $T$ and the KPZ height at time $t$ reads

\begin{eqnarray}
H &=& \log(Z(V)/\overline{Z})\,, \label{eq:H} \\
t &=& \frac {2L}{T^4} \label{eq:time}
\end{eqnarray} 
where  $\overline{Z}$ is the disorder average partition function.
We are interested in the distribution $P(H,t)$ over the disorder.

\begin{figure}
\includegraphics[width = 0.95\linewidth]{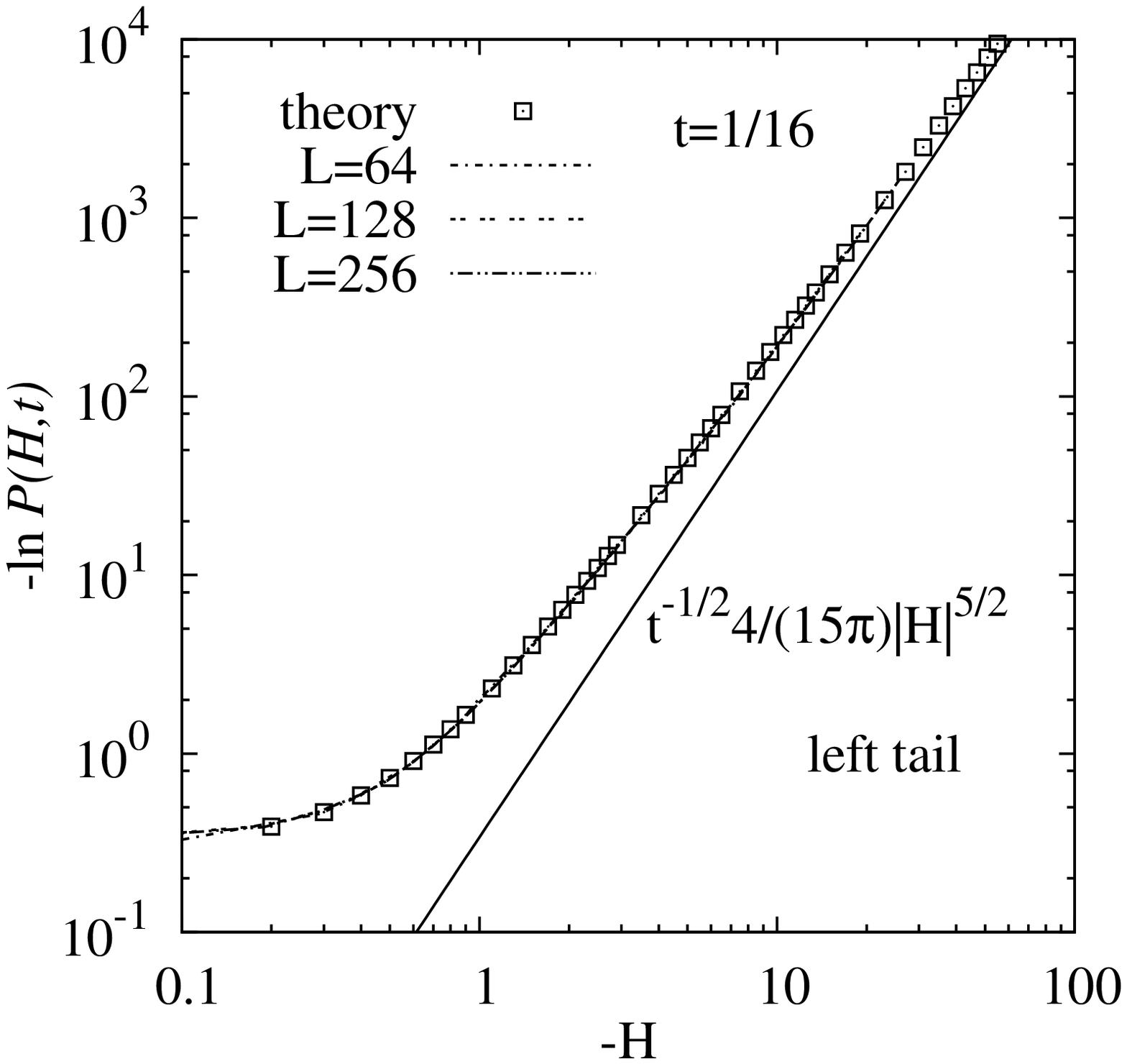}

\includegraphics[width = 0.95\linewidth]{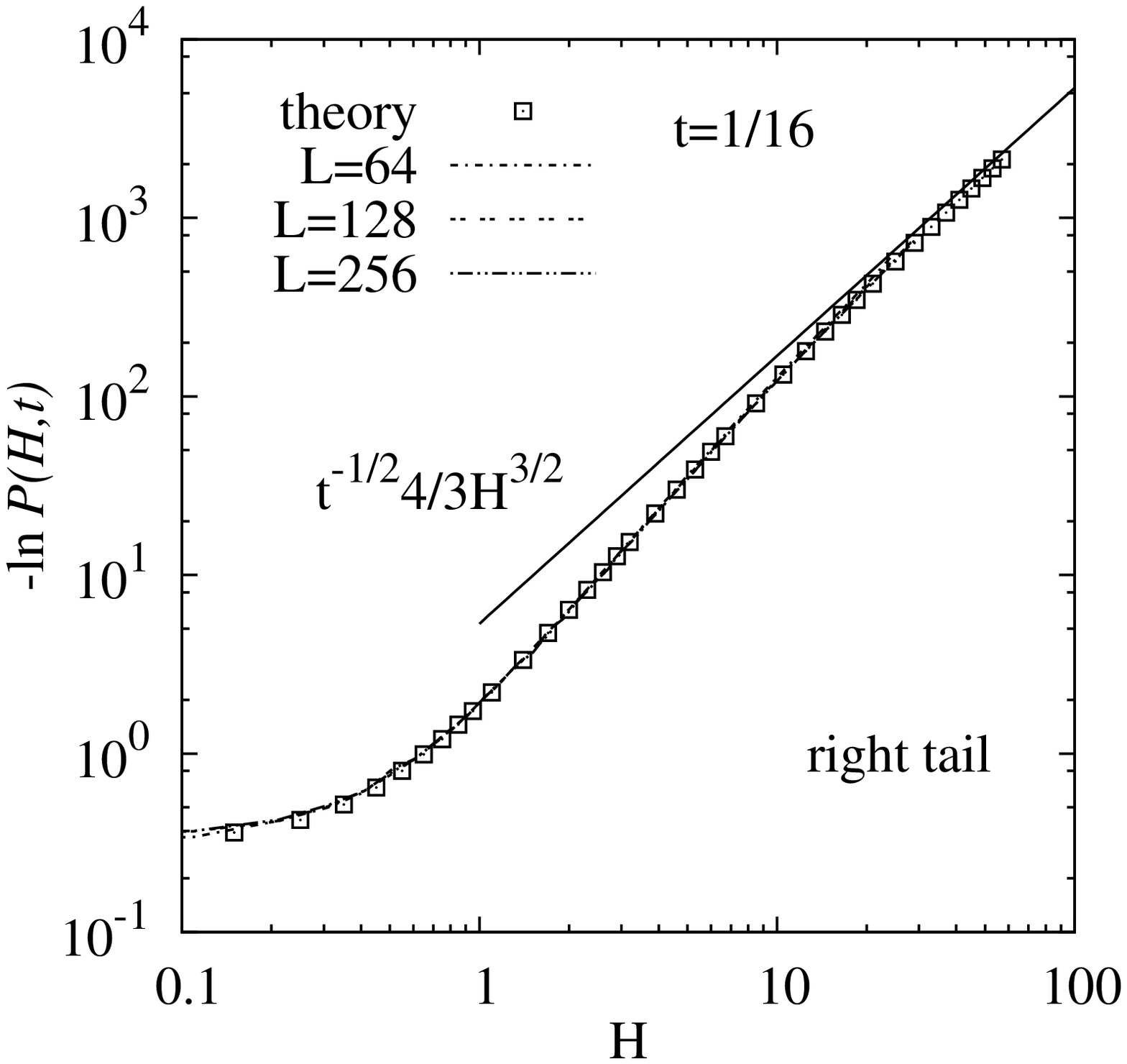}
\caption{{\bf Top}: blow up of the left tail of the data shown in Fig. \ref{Fig1} compared to the analytical prediction given in the first line of Eq.~(\ref{asympt3}). {\bf Bottom}: blow up of the right tail data shown in Fig. \ref{Fig1} compared to the analytical prediction given in the third line of Eq.~(\ref{asympt3}). 
\label{fig:P_H_0625}}
\end{figure}

{\em The importance sampling algorithm.} In principle one could obtain an estimate of 
$P(H,t)$ numerically from \emph{direct
sampling}: One generates many disorder realisation (say $\sim10^6$). For each realisation $Z(V)$ is computed. Then
 $\overline{Z}$ is estimated by averaging over all samples, and the
distribution is the histogram  of the values of $H$ according to Eq.~(\ref{eq:H}).
Nevertheless, this limits
the smallest probabilities which can be resolved, e.g., $10^{-6}$.

Therefore, we follow here a different approach.
To estimate $P(H,t)$ for a much larger range, where probabilities (or corresponding densities) 
smaller than, e.g., $10^{-1000}$ may appear, we will use
a more powerful approach, called  \emph{importance sampling} as discussed in Ref.~\cite{align2002,largest-2011}. This approach  has been succesfully applied in many cases to obtain the tails of distributions arising in equilibrium and
non-equilibrium situations, e.g.,   number of components of
Erd\H{o}s-R\'enyi (ER) random graphs \cite{rare-graphs2004},
the partition function of Potts models \cite{partition2005},
ground-state energies of directed polymers in random media \cite{monthus2006},
 the distribution of free energies of RNA secondary
structures \cite{rnaFreeDistr2010}, some large-deviation properties
of random matrices \cite{driscoll2007,saito2010}, the distribution
of endpoints of fractional Brownian motion with absorbing 
boundaries \cite{fBm_MC2013},
the distribution of work performed by an Ising system 
\cite{work_ising2014}, or the distributions of area and perimeter
of random convex hulls \cite{convex_hull2015,convex_hull_multiple2016}.

To keep the paper self-contained we now briefly outline the method.
Note that the approach has already been applied, in a slight variant,
to directed polymers in disordered media, at zero temperature 
\cite{monthus2006}.
The basic idea is to sample the different disorder realisations with 
an additional exponential bias $\exp(-\theta H(V))$ with $\theta$  as 
adjustable parameter. Note that if $\theta>0$ the configurations with a negative $H$ become more likely, conversely
for $\theta<0$ the configurations with a positive $H$ are favoured.   
A standard Markov-chain Monte Carlo simulation is then used to sample the biased configurations 
 \cite{newman1999,landau2000}. At each time step a new disorder realisation $V^*$ is proposed by replacing on the current realisation $V$ 
a certain fraction $r$ of the random numbers $V[x][y]$ by new Gaussian numbers.
The new disorder realisation is then accepted with the Metropolis-Hastings probability

\begin{equation}
p_{\rm Met} = \min\left\{1,e^{-\theta\left[H(V^*)-H(V)\right]}\right\}
\end{equation}
otherwise the old configuration is kept \cite{metropolis1953}.
Note that the average partition function $\overline{Z}$ appearing
in the definition of $H$ (\ref{eq:H})
drops out of the Metropolis probability, i.e., it is not needed here.
 By construction,
the algorithm fulfils detailed balance. Clearly the algorithm is also
ergodic, since within a sufficient number of steps, 
each possible realisation may be constructed. Thus,
in the limit of infinitely long Markov chains,
the distribution of biased disorder realisations will follow the probability

\begin{equation}
q_\theta(V) = \frac{1}{Q(\theta)} P_{\text{dis}}(V)e^{-\theta H(V)}\,, \label{eq:qT}
\end{equation}
where $ P_{\text{dis}}(V)$ is the original disorder distribution (here a simple product of independent Gaussians) and $Q(\theta)= \sum_V P_{\text{dis}}(V)e^{-\theta H(V)} $ is the normalization factor. Note that $Q(\theta)$ also depends
on $L$ and $T$, which we omit here in the notation for brevity. 
$Q(\theta)$ is generally unknown but can be determined, see below.
Thus the output of this Markov chain allows to construct a biased histogram $ P_\theta(H,t)$. In order to get the correct histogram $P(H,t)$ one should re-weight the obtained result:

\begin{equation}
 P(H,t) =  e^{\theta H} Q(\theta,t) P_\theta(H) \quad .
\label{eq:rescaling}
\end{equation}
Hence, the target distribution $P(H,t)$ can be estimated, up to a normalisation
constant $Q(\theta)$.  For each value of the
parameter $\theta$, 
a specific range of the distribution $P(H,t)$ will be sampled:
using a positive (respectively negative) parameter allows to sample the region of
a distribution at the left (respectively at the right) of its peak. 

{\em Technical details.}  To sample a wide range of values of $H$, one chooses a
suitable set of parameters $\{\theta_{-N_{n}},\theta_{-N_{n}+1},\ldots,
\theta_{N_{p}-1},\theta_{N_{p}}\}$,
$N_{n}$ and $N_{p}$ being the number of negative
and positive parameters, to access the large deviation regimes (left and right). 
The normalisation constants $Q(\theta)$ are obtained 
by first computing the histogram using direct sampling, which is well normalised and corresponds to $\theta=0$.
Then for $\theta_{+1}$, one matches the right part of the biased histogram with the left tail of the unbiased one and for $\theta_{-1}$, one matches the left part of the biased histogram with the right tail of the unbiased one. Similarly one iterates for the other values of $\theta$ and the corresponding \emph{relative} normalisation
constants  can be obtained.

The main drawback of our method is that as for any Markov-chain Monte Carlo simulation, it has to be
equilibrated and this may take a large number of steps. To speed the
simulation up, \emph{parallel tempering}  was used \cite{hukushima1996}. Here, a parallel
implementation using the \emph{Message Passing Interface (MPI)} was
applied, such that each computing core was responsible in parallel 
for an independent realisation
$V_{i}(s)$ at a given $\theta_i$. 
After 1000 Monte Carlo steps, one parallel-tempering
sweep was performed and the parameters  $\theta_i$ and $\theta_{i+1}$
were exchanged between two computing cores. The parameter $r$ is fixed
with criterion that the empirical
 acceptance rate of the parallel-tempering
exchange step is about 0.5  for all pairs of neigboring $\theta_i$.
A pedagogical explanation and examples of 
this sampling  procedure can be found in 
Ref.~\cite{align_book}.

\section{Results \label{sec:results}}

We have performed extensive numerical simulations 
\cite{practical_guide2015}
for polymer lengths
$L=64, 128$ and 256 and considered three different times corresponding  to short times $t \ll 1$ ($t=1/16$, $1/4$) and 
 (quite) large times  ($t=32$). In the numerical simulations the temperatures $T$
were chosen according to Eq.~(\ref{eq:time}). 

For each set of values $L$ and $T$, the numbers $N_{n}$ and $N_{p}$
and the values of parameters $\{\theta_{-N_{n}},\ldots,\theta_{N_{p}}\}$
were determined from numerical experiments. For small sizes $L=64$ the
number $N_{n}+N_{p}$ of parameters was typically about 30 
with values, e.g.,
 $\theta\in [-0.5, -0.015 ]\cup [0.06, 0.5 ]$. For the largest size $L=256$
up to $N_{n}+N_{p} =117$ different parameter values in the range
$[-0.013, -0.2] \cup [ 0.3, 1] $ were used.  Depending on the value of 
$\theta$, the Markov-chain variation parameter $r$
 ranged between 3.6\% (large $|\theta|$, i.e., $\theta=-0.2$ and $\theta=1$) 
and 0.018\% (smallest $|\theta|$, i.e., $\theta=-0.013$ here).

We first study the distribution $P(H,t)$ computed with the importance 
sampling algorithm explained above for the short time 
$t=1/16$. The results are 
shown in Fig. \ref{Fig1}  for different lengths $L=64,128$ and 256
and we compare the numerical results with the 
analytical result given in Eq.~(\ref{eq:ST}). The agreement for negative $H$ is very accurate for all lengths, over
800 decades in probability. For positive $H$ slight deviations are visible, but they
become smaller with increasing the length $L$ of the polymer, indicating a convergence to the 
analytical results as well. The behaviour of the extreme left and right tails is also shown in Fig. \ref{fig:P_H_0625}. 

\begin{figure}
\begin{center}
\includegraphics[width=0.95\linewidth]{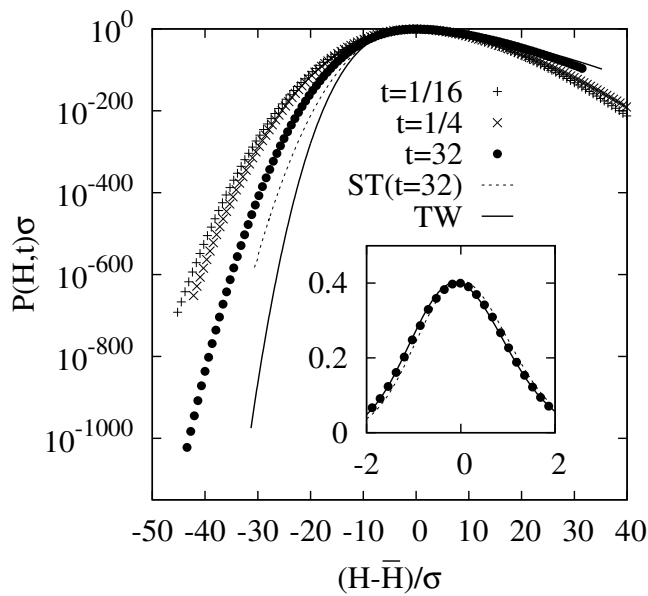}
\end{center}
\caption{Distribution of $P(H,t)$ for short ($t=1/16$), medium ($t=1/4$)
and longer time ($t=32$) for the longest length $L=256$. All data is normalized
to mean zero and variance one. The solid line shows the Tracy-Widom
distribution, the dashed line the short-time result given in Eq. (\ref{eq:ST}) with $t=32$. The inset magnifies the region of high
probability for the $t=32$ case and the two analytical results.
\label{fig:P_H_compare}}
\end{figure}

In Fig. \ref{fig:P_H_compare} the distributions $P(H,t)$ are shown for increasing
times $t=1/16$, $t=1/4$ and $t=32$ together with the Tracy-Widom and the short-time
distributions. Here we want to compare only the  distribution shapes and
therefore we have normalized all the curves  
to have mean zero and unit variance.
Regarding the relatively large time $t=32$ in the typical
region (Fig. \ref{fig:P_H_compare} inset) the numerical data clearly differ from the short time predictions and are closer to the Tracy-Widom distribution.
The right tail is very well described by the behaviour predicted in Eq. (\ref{3}) 
and Eq. (\ref{eq:phi+}) but the far left tail clearly differs from the Tracy-Widom tail.

\begin{figure}
\begin{center}
\includegraphics[width=0.95\linewidth]{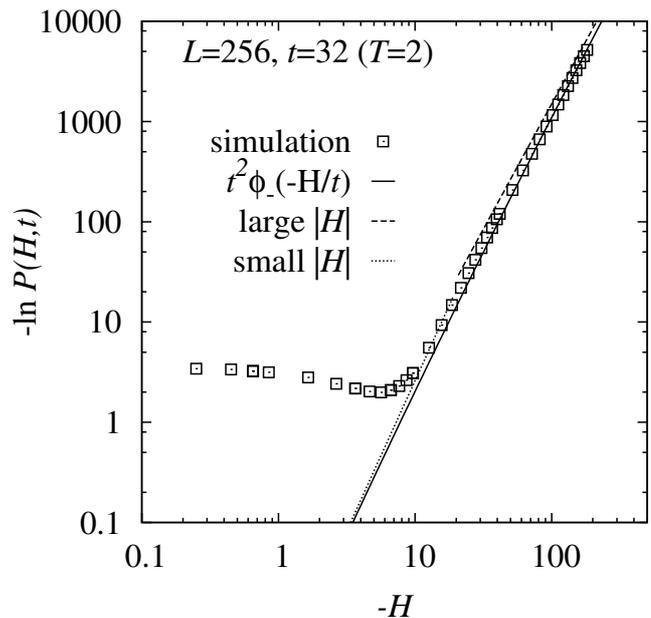}
\end{center}
\caption{Logarithm of the left tail of $P(H,t)$ for  longer time ($t=32$) 
and for the longest length $L=256$, shown in double-logarithmic scale.
The solid line shows the analytical prediction of Eq.~(\ref{eq:phi_m}). The broken
line shows the resulting limiting power-law: $|H|^3/(12t)$ for very large $H$, and  $\frac{4}{15 \pi} |H|^{5/2}/\sqrt{t}$  for moderate large $H$.
\label{fig:log_P_H_theorey_t32}}
\end{figure}

To investigate further the long-time behavior in the negative $H$ tail, 
we compare the result for $t=32$ directly with the analytic result in
Eq.~(\ref{eq:phi_m}). For better visibility, $-\ln(P(H,t))$ is shown in
Fig.~\ref{fig:log_P_H_theorey_t32} together with the analytic 
prediction of Eq.~(\ref{eq:phi_m}).
 For the largest values
of $-H$ accessible here, 
a convergence towards the power law $(-H)^{5/2}$ can be observed.
Note that the limiting $(-H)^3$ behavior for small values of $z=H/t$ is not visible
here. This is presumably because this regime is too close to the peak of the distribution.
Nevertheless, a small bending is visible in the log-log plot,
indicating an increase of the power towards $3$ for small values
of $-H$.

To summarize, a large-deviation sampling approach has been used
to measure the distribution $P(H,t)$ of heights for the KPZ equation
with a droplet initial condition. This was achieved using a lattice directed polymer 
model, whose free energy converges in the high temperature limit to the height of the
continuum KPZ equation. 
This allowed us to determine numerically the probability 
distribution of the height over a large range of values, allowing for a precise
comparison with the analytical predictions. We find that the agreement
with the short time large deviation function $\phi_{\rm short}(H)$ predicted by the theory \cite{leDoussal2016short} is
spectacular, even very far in the tails. Although we cannot strictly reach the 
large time limit, our intermediate time results are consistent with
both the $|H|^{5/2}$ (negative) and $H^{3/2}$ tails predicted by the theory \cite{sasorov2017}.
Our conclusion is that these far tails are mostly stable in time.

\begin{acknowledgments} 
AKH is grateful to the LPTMS for hosting and financially supporting
him for two months during his sabbatical visit July and September 2016.
The simulations were mostly performed at the HPC clusters HERO and CARL, both
        located at the University of Oldenburg (Germany) and funded by the DFG
        through its Major Research Instrumentation Programme
        (INST 184/108-1 FUGG and INST 184/157-1 FUGG) and the Ministry of
        Science and Culture (MWK) of the Lower Saxony State. This research was partially 
        supported by ANR grant ANR-17-CE30-0027-01 RaMaTraF.
\end{acknowledgments}

\bibliography{refs_grep.bib}

\end{document}